# Genus statistics for structure formation with topological defects

P. P. Avelino

*Department of Applied Mathematics
and Theoretical Physics
University of Cambridge
Silver Street, Cambridge CB3 9EW* \*

## Abstract

The main objective of this paper is to study the efficiency of genus statistics in differentiating between different models of structure formation. Some of the models studied are over-simplified but aim at reproducing some of the features of the structure seeded by topological defects. We consider accretion onto static point masses that could approximate accretion onto slow-moving cosmic string loops or other primordial point-like sources. Filamentary structures and wakes are also considered as an approximation to the structures seeded by slow-and-fast-moving long wiggly strings. Comparisons are also made with predictions of genus statistics for Gaussian fluctuations and with genus curves obtained from the CfA survey. A generic class of density models with wakes and filaments was seen to provide results comparable or better than Gaussian models for this suite of tests.

Subject headings: cosmology: large-scale structure of the universe: cosmic strings

## 1 Introduction

There are several models that aim at explaining the large scale structure we observe in the universe today. The two main paradigms for the origin of the seed perturbations are, on one hand, the inflationary scenario that produces Gaussian fluctuations and on the other hand, topological defects produced at phase transitions in the early universe that produce a non-Gaussian spectrum of density perturbations, specially on small scales. While, in general, inflationary scenarios require great fine tuning of parameters in order for the required perturbation amplitude to be produced by the present time, models where the structure is seeded by topological defects depend mainly on a single parameter which usually is the energy scale of the phase transition. It turns out that for cosmic strings, global monopoles and textures the energy scale is that of GUT phase transitions. In a way these models seem to explain the origin of the seed perturbations more naturally than inflationary models and so they deserve close attention. At present most work is done on inflationary models and, because they produce Gaussian fluctuations, all statistics are completely described by the power spectrum; other statistics that could help to distinguish between inflationary and non-inflationary models are often neglected. Because both models produce a nearly scale-invariant spectrum of fluctuations and because the uncertainties in the measurements are still large, the power spectrum is not the best statistic to distinguish between them. So, a higher order statistic is required. In this article we investigate the efficiency of genus statistics in distinguishing between different models of structure formation such as inflationary and cosmic string models and compare with observational genus curves obtained from the CfA survey. Some work on genus curves of isodensity contours for toy models of structure formation seeded by topological defects was previously done by Brandenberger, Kaplan & Ramsey (1993). By requiring that all the structures have the same size and mass they appear not to have properly taken into account the scaling solution. The size of the defect seeds increases with time (proportionally to the horizon) and the amplitude of the density perturbations induced by larger defects is smaller because they have less time to grow by gravitational instability. This effect is properly taken into account in the present paper and so we consider our work an improvement over that of Brandenberger et al. (1993). Robinson & Albrecht (1995) also performed a similar study with cosmic string wakes. Their cosmic string toy model consisted of a realization of a string power spectrum, where the phases of the Fourier modes were choosen at random, plus a single cosmic string wake. They concluded that the genus statistic is not a good discriminator between their model and a model without the wake included. As we will show in this paper the genus statistic is a good discriminator between our cosmic string toy models and random phase models with the same power spectrum at least in the linear regime. So, we conclude that their model underestimates the presence of sheet-like features in the density field and, therefore, in this regard it should not be considered a good cosmic string toy model.

---

\* Email: ppa1000 @ amtp.cam.ac.uk

This article is organized as follows. We start in section 2 by introducing the Zel'dovich aproximation, solving it for accretion onto static point masses, filaments and wakes assuming the dark matter to be cold. In the end of the section a modification to the Zel'dovich aproximation that accounts for the neutrino free streaming length is introduced when considering a hot dark matter model. In section 3 the genus statistic is described. Analytic results for Gaussian perturbations are given and topological measures of departures from Gaussianity are introduced. In section 4 we describe the way we produce the fluctuations for the models considered and relate that to the results of section 2. In section 5 the results are presented. The dependence of the genus curves on the type and number of topological defects present is demonstrated and error bars due to sample variance are introduced. The density probability distribution for several of the models studied in this article are given and the parametrization of the genus curve discussed. The final section is a discussion of the results.

In this article we shall assume that $\Omega = 1$ and $h = H_0/(100 \text{ Km/sec/Mpc})$.

## 2 The Zel'dovich approximation

### 2.1 Cold dark matter

We employ the Zel'dovich approximation (Zel'dovich 1970) to examine each model we use. We will assume the universe to be flat, with no cosmological constant and the dark matter to be cold. In the next section we will see how the Zel'dovich approximation can be modified in order to describe hot dark matter as well. The equations for the evolution of the scale factor are the Friedman and Raychaudhuri equation that in the matter are

$$\left(\frac{\dot{a}}{a}\right)^2 = \frac{8\pi G \rho_{\rm b}}{3}, \tag{1}$$

$$\frac{\ddot{a}}{a} = -\frac{4\pi G \rho_{\rm b}}{3}, \tag{2}$$

where $\rho_b$ is the background density. The position of the CDM particles is given by

$$\mathbf{r} = a(t)[\mathbf{q} + \vec{\psi}(\mathbf{q}, t)], \tag{3}$$

where $\mathbf{q}$ is the unperturbed comoving position of the particle and $\vec{\psi}$ is the comoving displacement vector of the particle. In the presence of a perturbing string seed, the cold dark matter particle will obey

$$\frac{d^2 \mathbf{r}}{dt^2} = \mathbf{F}_{\text{seed}} + \mathbf{F}_{\text{matter}}, \tag{4}$$

where the force $\mathbf{F}_{\text{matter}}$ due to the surrounding matter is given by

$$\mathbf{F}_{\text{matter}} = -\nabla \Phi, \tag{5}$$

with the gravitational potential $\Phi$ satisfying the Poisson equation

$$\nabla^2 \Phi = 4\pi G \rho_{\text{matter}} = 4\pi G \rho_b (1 + \delta). \tag{6}$$

The fact that mass is conserved is described by the continuity equation that can be written in comoving coordinates to first order in $\vec{\psi}$ as

$$\frac{\partial (\rho a^3)}{\partial t} + \nabla_\mathbf{q} \cdot (\rho a^3 \dot{\vec{\psi}}) = 0, \tag{7}$$

which can be integrated in the limit of small perturbations to give

$$\delta = \frac{\partial \rho}{\rho} \approx -\nabla_\mathbf{q} \cdot \vec{\psi}, \tag{8}$$

again to first order in $\vec{\psi}$. The Poisson equation can then be integrated to give

$$\nabla_\mathbf{r} \Phi = \frac{4\pi G \rho_b}{3} (\mathbf{r} - 3a \vec{\psi}_\parallel), \tag{9}$$

where $\vec{\psi}_\parallel$ and $\vec{\psi}_\perp$ are defined by

$$\vec{\psi} = \vec{\psi}_\parallel + \vec{\psi}_\perp, \tag{10}$$

$$\nabla_\mathbf{r} \times \vec{\psi}_\parallel = 0, \tag{11}$$

$$\nabla_{\mathbf{r}} \cdot \vec{\psi}_{\perp} = 0. \tag{12}$$

In the linear regime with ($|\vec{\psi}| << |\mathbf{q}|$) we can use (4), (5) and (9), to write the Zel'dovich approximation.

$$a \left( \frac{\partial^2}{\partial t^2} + 2\frac{\dot{a}}{a}\frac{\partial}{\partial t} + 3\frac{\ddot{a}}{a} \right) \vec{\psi}_{\parallel} = \mathbf{F}_{\text{seed}\parallel}, \tag{13}$$

$$a \left( \frac{\partial^2}{\partial t^2} + 2\frac{\dot{a}}{a}\frac{\partial}{\partial t} \right) \vec{\psi}_{\perp} = \mathbf{F}_{\text{seed}\perp}. \tag{14}$$

If the perturbations are irrotational to begin with, and if the source term is irrotational $\vec{\psi}_{\perp} = 0$ so that $\vec{\psi} = \vec{\psi}_{\parallel}$. Then we can write the Zel'dovich approximation as

$$\left( \frac{\partial^2}{\partial t^2} + 2\frac{\dot{a}}{a}\frac{\partial}{\partial t} + 3\frac{\ddot{a}}{a} \right) \vec{\psi} = \frac{1}{a}\mathbf{F}_{\text{seed}} = \mathbf{S}_{\text{seed}}. \tag{15}$$

If the seed is a point mass $\mathbf{S}_{\text{seed}}$ is given by

$$\mathbf{S}_{\text{pmass}} = -\frac{GM(\mathbf{q}-\mathbf{q_s})}{|\mathbf{q}-\mathbf{q_s}|^3 a^3}, \tag{16}$$

where $\mathbf{q_s}$ is the comoving position of the point mass. If the seed is a line of mass then $\mathbf{S}_{\text{seed}}$ is given by

$$\mathbf{S}_{\text{lmass}} = -\frac{2GM_L(\mathbf{q}-\mathbf{q_s})}{|\mathbf{q}-\mathbf{q_s}|^2 a^2}, \tag{17}$$

where $\mathbf{q_s}$ is the comoving position of the point in the line of mass nearer to $\mathbf{q}$ and $M_L$ is the mass per unit length of the line of mass. The Zel'dovich approximation can be solved using the Green's function method. For the case we are considering the Green's function is

$$G(t, t_0) = \frac{3t_0}{5} \left( \left(\frac{t}{t_0}\right)^{2/3} - \frac{t_0}{t} \right) \qquad t > t_0, \tag{18}$$

$$G(t, t_0) = 0 \qquad t < t_0. \tag{19}$$

Using it we find the solution of the Zel'dovich equation for the static point mass and line of mass

$$\vec{\psi}_{\text{pmass}}(\mathbf{x}, t) = \frac{3GM\mathbf{x}t_i^2}{2|\mathbf{x}|^3} \left( 1 - \frac{2}{5}\frac{t_i}{t} - \frac{3}{5}\left(\frac{t}{t_i}\right)^{2/3} \right), \tag{20}$$

$$\vec{\psi}_{\text{filament}}(\mathbf{x}, t) = -\frac{6GM_L\mathbf{x}t_i^2}{5|\mathbf{x}|^2} \left( \left(\frac{t}{t_i}\right)^{2/3} \ln\left(\frac{t}{t_i}\right) - \frac{3}{5}\left(\frac{t}{t_i}\right)^{2/3} + \frac{3}{5}\frac{t_i}{t} \right). \tag{21}$$

The formation of wakes can be modeled given an initial velocity of the form

$$\dot{\vec{\psi}}(\mathbf{x}, t_i) = u_i \epsilon(\mathbf{q}), \tag{22}$$

where

$$\epsilon(\mathbf{q}) = -1, \quad \mathbf{x} \cdot \mathbf{q} > 0, \tag{23}$$

$$\epsilon(\mathbf{q}) = 1, \quad \mathbf{x} \cdot \mathbf{q} < 0. \tag{24}$$

This is given to the particles in cosmic string models because of the conical spacetime with $u_i = 8\pi G\mu v_s$, where $v_s$ is the string velocity. The solution of the Zel'dovich equation in this case is

$$\vec{\psi}(\mathbf{x}, t) = \frac{3}{5} u_i t_i \epsilon(\mathbf{q}) \left( \left(\frac{t}{t_i}\right)^{2/3} - \frac{t_i}{t} \right). \tag{25}$$

## 2.1 Hot dark matter

If the dark matter is hot, then small-scale perturbations with wavelength smaller than the neutrino (or other hot dark matter candidate) free streaming length $\lambda_{FS}$ will be erased. In order to properly describe the formation of structure in the context of a hot dark matter model one would need to solve the Boltzmann equation for the dark matter particles. However, it was shown by Perivolaropolous, Brandenberger & Stebbins (1990) in studying the clustering of neutrinos in cosmic string induced wakes, that most of the results can be described correctly using a naive modification of the Zel'dovich approximation. This modification is based on the fact that on average the dark matter particles will only start to collapse when the comoving free-streaming length has fallen below $|\mathbf{q}|$. So we modify the Zel'dovich approximation in the context of a hot dark matter model by setting $|\vec{\psi}| = 0$ on scales $|\mathbf{q}| < \lambda_{FS}$ and evolving scales $|\mathbf{q}| > \lambda_{FS}$ as for the cold dark matter case. We consider a hot dark matter model with two neutrino species with sufficient mass to make $\Omega_\nu = 1$. In particular we consider these two species to have the same mass $m_\nu = 46$ eV.

The comoving distance hot particles can move since $t \geq t_{\text{eq}}$ is

$$\lambda_{FS} = 3v(t)tz(t), \tag{26}$$

where $v(t)$ is the thermal velocity of the hot particles and $z(t)$ is the redshift. The free streaming length grows in the radiation era attaining a maximum at $t_{eq}$ and then decreases proportionally to $t^{-1/3}$ in the matter era. The maximal free streaming length is

$$\lambda_{FS} = 3 v_{eq} t_{eq} z_{eq}, \tag{27}$$

where

$$v_{eq} = T_\nu^{eq}/m_\nu \approx 0.09, \tag{28}$$

so that the maximum free streaming length is given by $\lambda_{FS}^{eq} \approx 3 h^{-2} \text{Mpc}$.

## 3 Genus statistics

### 3.1 Definition of genus

To measure the topology of isodensity contours we will use the Gauss-Bonnet theorem which relates the integrated Gaussian curvature (a local property) of a surface with the genus (a global property) of that surface. The Gaussian curvature of a two-dimensional surface at a particular point is

$$K = \frac{1}{a_1 a_2}, \tag{29}$$

where $a_1$ and $a_2$ are the two principal radii of curvature at that point. A surface has a positive or negative Gaussian curvature respectively if the two radii of curvature point in the same or in opposite directions. For example, a sphere has a positive radii of curvature given by $K = 1/r^2$ where $r$ is the radius of the sphere. A cylinder has $K = 0$ because one of the radii of curvature is infinite. Saddle points have negative curvature because $a_1$ and $a_2$ have opposite signs. The Gauss-Bonnet theorem relates the integral of the Gaussian curvature over the surface with the genus in the following way

$$\int K dA = 4\pi(1 - g), \tag{30}$$

where $g$ is the genus of the surface and $dA$ a surface element. The genus measures the number of closed curves that may be drawn on a surface without separating it. It can also be defined as

$$g = \text{number of compact regions} - \text{number of holes} + 1. \tag{31}$$

A curved surface may be approximated by a network of polygonal faces. When we use such a network to compute the genus we find that

$$\sum D_i = 4\pi(1-g), \qquad (32)$$

where $D_i = 2\pi - \sum V_i$ is the angle deficit at a vertex in radians. In this case the curvature is effectively compressed into delta functions at the vertex. We used (32) to construct a numerical alghoritm in C in order to determine the genus of an isodensity surface applying the method sugested by Gott, Mellot & Dickinson (1986). The isodensity surface is constructed by binning the density onto a cubic lattice and identifying pixels with density above and below a certain threshold ($\delta_c$). Our program to compute the genus was tested against analytic results for Gaussian perturbations and also against well known topological configurations for which we knew the genus beforehand (e.g. a network of isolated cubes).

### 3.2 Window function

The data is smoothed with a Gaussian window function

$$w(r) = \frac{1}{\pi^{3/2} \lambda_e^3} e^{-\frac{r^2}{\lambda_e^2}}, \qquad (33)$$

where $\lambda_e$ is the smoothing scale. This definition implies that the smoothing length is greater by a factor of $\sqrt{2}$ than the usual width of a Gaussian. The smoothing scales are always greater than the average interparticle spacing but not too large as to erase all the relevant features in the density map. The smoothing scales considered in this article were 6, 8, 10, 12, 16 and $20h^{-1}$Mpc in order to make a direct comparison with the results of Vogeley et al. (1994) for the CfA survey.

### 3.3 Genus for Gaussian random fields

A comparison with inflationary models predictions for the genus curve is essential if one has to decide which model describes better the kind of large scale structures we observe in the universe today. The genus curves of random fields are well studied and some analytic formulae have been derived (Bardeen et al. 1986; Hamilton, Gott & Weinberg 1986). The genus per unit volume is given by

$$g_s(\nu) = N(1-\nu^2)e^{-\nu^2/2}, \qquad (34)$$

where $\nu$ is the number of standard deviations above or below the mean density contour. The amplitude $N$ depends on the power spectrum of density fluctuations $P(k)$ as

$$N = \frac{1}{4\pi^2}\left(\frac{\langle k^2 \rangle}{3}\right)^{3/2}, \tag{35}$$

where

$$\langle k^2 \rangle = \frac{\int k^2 P(k) d^3k}{\int k^2 d^3k}. \tag{36}$$

If we smooth the structure on a scale $\lambda_e$ the power spectrum becomes

$$P'(k) = P(k)e^{-k^2\lambda_e^2/2}, \tag{37}$$

and if the power spectrum is of the form $P(k) \propto k^n$ then $N$ is given by

$$N = \frac{1}{(2\pi)^2 \lambda_e^3}\left(\frac{3+n}{3}\right)^{3/2}. \tag{38}$$

To compute the genus curve we must multiply (34) by the volume of the grid. In fig. 1 we plotted a random phase genus curve obtained for a $P(k) \propto k$ power spectrum obtained respectivelly analytically, using (38) and (34), and numerically using our program to compute the genus. The two curves are almost identical as we should expect if the program to compute the genus is working properly. Small differences between the two curves can be atributed mainly to the choice of periodic boundary conditions. Smaller differences are also due to the sample variance (the volume of the box is not infinite) and to small numerical imprecisions. The figure is symmetric with respect to the vertical axis which puts in evidence the topological equivalence of positive and negative linear density perturbations for random phase models of structure formation. The shape of the random phase genus curve is independent of the power spectrum. For $|\nu| < 1$ the genus is always positive, the surface has more holes than compact regions and so the surface is "spongelike". For $|\nu| > 1$ the genus is negative and the surface has a lot of independent compact regions. For non-random phase distributions the genus curve will be, in general, asymmetric because the topological symmetry between high and low density regions is not expected in most cases.

### 3.4 Genus metastatistics

To quantify departures from the random phase curve, Vogeley et al. (1994) used genus related statistics like the amplitude, the width, and the shift of the genus curve. The amplitude was defined as the amplitude of the best fit random phase curve. That was done by determining the amplitude of the genus curve that minimizes $\chi^2$. The reason for the definition was that they wanted to compare the observational data from the CfA survey with predictions from Gaussian models; So, it seemed appropriate. Although, we want to make a comparison of present observations with non-Gaussian models we will retain the same definition in order to directly compare our results with those of Vogeley et al. (1994).

The width of the genus peak $W_\nu$ was defined as the difference between the zero crossings of the genus curve which for random phases is equal to $W_\nu = \nu_+ - \nu_- = 2$ because, as we have seen before the genus for random phases is positive over the range $-1 < \nu < 1$ and negative elsewhere. This change of the genus sign is believed to coincide with the percolation thresholds for random-phase perturbations. In the range $-1 < \nu < 1$ both high ($\delta > \delta_c$) and low ($\delta < \delta_c$) density phases percolate, while for $\nu > 1$ only the low density phase percolates and for $\nu < -1$ only the high density phase percolates.

The last statistic they used was the shift $\Delta \nu$ of the genus curve which was quantified in the following form

$$\Delta \nu = \frac{\int_{-1}^{1} \nu G(\nu)_{\text{obs}} d\nu}{\int_{-1}^{1} G(\nu)_{\text{fit}} d\nu}, \tag{39}$$

where $G(\nu)_{\text{obs}}$ is the measured genus curve and $G(\nu)_{\text{fit}}$ is the best fit random phase cuve. A positive value of $\Delta \nu$ indicates a density distribution that is more 'meatball-like' (isolated cluster models) than random phase, while a negative value of $\Delta \nu$ is characteristic of a 'bubble-like' topology ('swiss chesse' topology).

### 3.5 Parametrization of the genus curve

For a Gaussian density field the volume fraction in the high density region is given by

$$f(\nu > \nu_0) = \frac{1}{\sqrt{2\pi}} \int_{\nu_0}^{\infty} e^{-t^2/2} dt. \tag{40}$$

Vogeley et al. (1994) did not compute the mean and standard deviation $\nu$ of the density distribution and express the genus curve as a function of $\nu$. Instead they determined the genus curve as a function of the volume fraction in the high density region and used (40) to parametrize $f$ as a function of $\nu$. Although for Gaussian perturbations these two ways of calculating the genus curves would give similar results, for some of the models we investigate in this article the genus curves are quite different due to their non-Gaussianity.

### 3.6 Smoothing of the genus curve

The genus curves were smoothed using a very simple procedure known as three-point boxcar smoothing (see for example Vogeley et al. 1994) that was shown to give better estimates of the true genus curve for Gaussian random phase models. It consists in determining the genus as

$$G'(\nu_i) = \frac{1}{3}(G(\nu_{i-1}) + G(\nu_i) + G(\nu_{i+1})), \tag{41}$$

where $\nu_{i+1} = \nu_i + 0.1$.

## 4 Models and observations

### 4.1 Toy models

The toy models we study in this paper are simplified cosmic string models. It is a well-known fact that the gravitational effect of a slow-moving small loop can be well approximated by that of a static point mass, thus generating spherical accretion (see for example Vilenkin & Shellard 1994). Also a slow-moving wiggly string can be approximated by a line of mass, thus generating filamentary structures, while a fast moving string generates a wake. It is also known that cosmic string networks after the friction dominated era rapidly attain a scaling solution where the average properties of the network (such as the average number of defects and the average correlation length) remain the same at all times when scaled to the horizon size (Bennett & Bouchet 1990; Allen & Shellard 1990, Albrecht & Turok 1989). This means that although fluctuations laid down at later times are smaller in amplitude because they have less time to grow by gravitational instability, they will have a larger wavelength (in proportion to the horizon size). These facts are essential ingredients of the toy models we consider in this paper.

Here we consider several toy models of cosmic string structure formation without taking into consideration the detailed properties of cosmic string networks known from numerical simulations. These models are respectively a network of fast-moving strings, a network of slow-moving wiggly strings and a network of slow-moving small loops. Both the filament and wake models should give a rough approximation to the structures seeded by cosmic strings. Although the network of slow moving small loops cannot be considered a realistic cosmic string toy model because loops of cosmic string produced by a cosmic string network are born with relativistic velocities and are in much higher number than what is considered in this article, there are scenarios which this toy model does approximate, notably those with loop nucleation during inflation. The kind of shapes we investigate also appear in other defect-seeded structure formation models like those seeded by global monopoles or global textures. Although we want primarily to test if the genus statistic is a good discriminator between different models of structure formation (specially between different non-Gaussian models) we also want to see if some of the features of these simplified models match current observations. To properly test cosmic string models of structure formation one would need to go beyond this simplified models and perform large-scale network simulations (Avelino & Shellard 1995).

In fig 2 we plotted the isodensity contours for some of the CDM toy models considered. We can see mainly filamentary, wake-like, and spherical structures respectively in the 5 filament, 5 wake and 25 sphere models. Other kinds of shapes can be obtained due to superposition of density perturbations generated by several defects. In the sphere model there are more small spheres than big ones because denser objects are generated later in this model when there are less defects inside the box. In the filament and wake models we can see more smaller wakes and filaments than larger ones. This is due to the fact that smaller objects are seeded earlier when there are more defects inside the box. These give rise to larger density perturbations and so to thicker objects in the density contour plot.

### 4.2 The CfA survey

Vogeley et al. (1994) studied the topology of large scale structure in the CfA Redshift Survey. This survey includes $\approx$ 12000 galaxies with limit magnitude $m_b \leq 15.5$ and allowed for the computation of the topology on smoothing scales from 6 to $20h^{-1}$Mpc.

They used genus statistics to test several variants of the cold dark matter cosmogony (CDM). All of them failed to match the observations to a high confidence level ($> 97\%$) (Vogeley et al. 1994), even when evolution of the perturbations into the non-linear regime through the use of N-body codes was taken into account. This is a good motivation for our work based on non-Gaussian models of structure formation. Our results for the cosmic string toy models of structure formation are such that direct comparisons with the CfA genus curves are possible.

### 4.3 Generation of fluctuations

We apply the genus statistics to smoothing scales between $6h^{-1}$Mpc and $20h^{-1}$Mpc that are in the linear or mildly non-linear regime by the present time. In this paper we will only consider linear theory, in the form of the Zel'dovich approximation, to evolve the perturbations in the matter era. Matter accretion during the radiation era was not considered in the present paper. For linear perturbations the genus curve is isomorphic to the 'initial' genus curve which implies that the genus curve as a function of the number of standard deviations from mean density does not change with time if no additional defects enter the box. The effect of non-linear evolution on the genus curves is work in progress at the present moment (Avelino & Canavezes 1995). However, we expect this to be small on scales greater than $8h^{-1}$Mpc. We produce the density fluctuations in a comoving box whose size is chosen so that the genus curves obtained can be directly compared with those obtained from the CfA survey. We fix the scaling solution by fixing the length and number of defects of each type per horizon volume. These were put in the box with random positions and orientations. The number of defects per horizon volume is $N \pm \sqrt{N}$ and the wavelength of the perturbations induced by them to be initially (at $t_{\text{eq}}$) $\lambda_{\text{eq}}$ and $\frac{2}{3}\lambda_{\text{eq}} \times \frac{2}{3}\lambda_{\text{eq}}$ for the filaments and wakes respectively ($\lambda_{\text{eq}} = ct_{\text{eq}}$). Structures seeded at later times will have a larger wavelength due to the scaling solution. These structures were produced by displacing the particles according to the symmetry point, axis and plane and with a dependence on the spatial coordinate given by each of the solutions of the Zel'dovich equation. In appendix 1 we give a more detailed description of how the fluctuations are produced for each toy model. In this article we assume the biasing parameter does not depend on the scale so that we can directly compare the genus curves obtained from the CfA survey with those obtained for the toy models considered in this paper (at least for scales which are in the linear regime) which deal only with the distribution of the dark matter .

### 4.4 Compensation

Arbitrary energy momentum perturbations are not possible in a Robertson-Walker spacetime (Trashen 1985; Trashen, Turok & Brandenberger 1986; Veeraghavan & Stebbins 1990). When the strings are formed in the early universe the excess energy and linear momentum carried by the string is compensated by an equal deficit in the background radiation. The fact that energy and momentum must be conserved is well expressed by the following integral constraints

$$\int_V \delta\rho dV \approx 0, \qquad (42)$$

$$\int_V \delta\rho \mathbf{x} dV \approx 0, \qquad (43)$$

where V is a volume much greater than the horizon volume. Because of the way in which we generate the perturbations, only positive density perturbations would be allowed if compensation was not taken into account. Although it is difficult to mimic the detailed effects of compensation in the genus curve we can at least, as a first approximation, correct the genus curve for the average shift of the peak due to not including compensation by doing

$$\delta_{\text{new}} = \delta_{\text{old}} - \delta_{\text{c}}, \qquad (44)$$

such that $\langle \delta_{\text{new}} \rangle = 0$ inside the box.

## 5  Results

### 5.1 Toy models and CDM

In fig 3 we show a comparison of the density probability distribution for 3 of the models studied in the context of cold dark matter, averaged over 10 simulations. These were the 5 filament model, the 5 wake model and the 25 sphere model.

We can see from the pictures that for $\lambda_e = 6h^{-1}$Mpc the density distribution is very non-Gaussian and that it will approach Gaussianity as we go to larger smoothing scales.

We can also see that the largest departures from Gaussianity occur for the 5 filament model while for wakes the density probability distribution is very nearly Gaussian even at small scales. If all the models had the same number of objects per unit volume we would expect the sphere model to exhibit the largest departures from Gaussianity, while the wake model should be the most Gaussian. The reason for this is that while a point-like overdensity can only be inside or outside a chosen volume, line-like perturbations and planar perturbations can be partially inside several volumes (a planar perturbation with size l × l being able to touch more volumes than a line-like perturbation of size l). Consequently, the number of pieces contained in a chosen smoothing volume will be maximized for the wake model, making this model more Gaussian, and minimized for sphere model making it less Gaussian. The reason why our particular sphere model is more Gaussian than the filament model is that there are more objects per unit volume in the sphere toy model than in the filament toy model.

The large departure from Gaussianity, especially on small scales, makes the genus curve very sensitive to the parametrization so that different genus curves are obtained if the genus is expressed directly as a function of the number of standard deviations from mean density or if we use the volume fraction parametrization for $\nu$. A comparison of two genus curves for the 5 filament model is shown in fig 4 in order to illustrate the effect of the parametrization by volume fraction. The genus curves are considerably different (the one parametrized by volume fraction being more like random phase), but as we have seen this effect should be most exaggerated for the 5 filament model at the smallest scale considered ($\lambda_e = 6h^{-1}$Mpc). For other toy models and smoothing scales the dependence of the genus curve on the parametrization will not be so considerable. However, for small smoothing lengths some information about the density probability distribution is lost. In this paper we shall use the volume fraction parametrization (unless it is said otherwise) used by Vogeley et al. (1994) when analysing the CfA observational data.

In fig 5 and fig 6 we show a comparison of the genus curves, for our 6 CDM models of structure formation, with the predictions from the standard inflationary CDM model, as well as the observations taken from the CfA survey. Error-bars due to sample variance were not included for the sake of clarity. The sizes of the boxes for the toy models studied in this article were chosen such that a direct comparison between their genus curves and CfA genus curves is possible for each of the smoothing scales considered. It is readily apparent from the graphs that from the models studied the one that best fits the observational data, at least from a topological point of view, is the 5 wake toy model. The amplitude of the genus curves seems to be only marginally larger than the CfA genus curves. Also, the width of the genus curves seems to mimic observations much better than the standard CDM model. The 10 wake model also does marginally better than standard CDM. The amplitude of the genus curves is always smaller than that predicted by the standard CDM scenario and in better agreement with observations.

The standard CDM model gives too large amplitude of the genus curve, especially on small scales and it fails to match other features of the observed genus curves. For example, the standard CDM model gives $W_\nu \approx 2$ and although sample variance allows some fluctuations around this value it is not enough to explain genus peak widths as large as 2.5 or 2.6 with a very large confidence level ($> 90\%$) (Vogeley et al. 1994). It is possible to have other random phase models with a smaller amplitude of the genus curve, as in open models or models with a non-zero cosmological constant. What happens in this case is that the growth of density perturbations in the linear regime slows down while the non-linear growth of perturbations proceeds in the normal way. So, for the same normalization of the spectrum of density perturbations non-linear perturbations are more non-linear in these models. Non-linearities introduce correlations between phases of different Fourier modes, decreasing the number of independent structures and consequently reducing the genus amplitude. However, as found by Vogeley et al. (1994) the problem of matching the other statistics, especially the width of the genus peak, remains unsolved.

The 5 and 10 filament models do better than spheres and marginally worse than standard CDM. The genus amplitude is higher than for standard CDM. However, these models provide a better fitting to the width of the observed genus curves than the standard CDM inflationary scenario. The sphere model is clearly ruled out. The amplitude of the genus curves is too large and it fails to match the shape of the observational genus curves.

Also apparent from the graphs is that for most of the smoothing scales considered, the genus amplitude is an increasing function of the number of defects. This should be expected because we are increasing the number of structures present inside the box.

In fig 7 we plotted the genus curves obtained for the best model (the 5 wake model) now with the error bars properly included. The line represents the average genus curve among 10 realizations of the model. The error bars are one sigma error bars over these realizations. The asymmetry of the genus curves is visible, as we would expect from almost any non-Gaussian model of structure formation. However, the parametrization by volume fraction makes the genus curves look less asymmetric, especially on small scales, making the curves more like random-phase models.

In fig 8 we show a statistical comparison of the genus curves for the wake model with the predictions from the CfA survey. The genus amplitude for this model is marginally larger than the observed genus amplitude for some the scales considered, although it does much better than standard CDM at matching the amplitude of the CfA genus curves. Again, the error bars we see in the plots are one sigma error bars of 10 simulations. The shift of the genus curves does not seem to be in agreement with observations for most of the scales considered. The width of the genus peak seems to be consistent with observations from the CfA survey for all of the scales considered but in this case the error bars due to sample variance are very large so that a very wide range of genus peak widths are possible. This toy model fits the observations better than any random phase model tried by Vogeley et al. (1994).

### 5.2 Toymodels and HDM

In fig 9 we show a comparison of the density probability distribution for 3 of the models studied as a function of the smoothing length and averaged over 10 simulations. These were the 5 filament model, the 5 wake model and the 25 sphere model with hot dark matter. These models are more Gaussian than the corresponding models with CDM and consequently the dependence of the genus curves on the parametrization is smaller in this case. For the wake model the density probability distribution is nearly Gaussian even for $\lambda_e = 6h^{-1}$Mpc.

In fig 10 and fig 11 we show a comparison of the genus curves for 6 models of structure formation we studied in the context of hot dark matter with the predictions from the standard CDM model and observations taken from the CfA survey. The amplitude of the genus curves is smaller with HDM than with CDM, especially on small scales, because an adittional smoothing was introduced due to the free-streaming of the neutrinos. The models that best mimic the observational results are again the wake models (particularly the 10 wake model).

In fig 12 we can see the genus curves obtained for the best model (the 10 wake model) now with the error bars properly included. These seem to be more like random-phase than the genus curves obtained for the 5 wake model. The line represents the average genus curve among 10 realizations of the model and the error bars are one sigma error bars.

Figure 11 shows a statistical comparison of the genus curves for the 10 wake model with the predictions from the CfA survey. The genus amplitude for this model seems to be consistent with the observed genus amplitude for most of the scales considered. Again, the error bars we see in the plots are one sigma error bars of 10 simulations. The shift of the genus curves does not seem to be in agreement with observations for most of the scales considered. The width of the genus peak seems to be consistent with observations from the CfA survey although in this case the error bars due to sample variance are very large. This toy model also fits the observations better than any random phase model tried by Vogeley et al. (1994).

## 6  Discussion

We can conclude from the results presented in this article that the genus statistic is a good discriminator between Gaussian and non-Gaussian models of structure formation. In addition to this, it was shown to be a good statistic to distinguish between different toy models of structure formation, sensitive to the shape, number of structures seeded, and dark matter type. We also showed that at least for some of the models considered (in particular the wake models) that there is a better agreement with the observations than that verified for random phase models of structure formation. It is necessary to use string network simulations in combination with numerical codes that generate and evolve the density perturbations seeded by such networks in order to properly test the cosmic string model for structure formation. However, we have shown that there are some features on the observed genus curves (e.g. genus peak width considerably larger than 2 and genus amplitude smaller than standard CDM), that cannot be easily matched by random phase models of structure formation, but which are matched by some toy models considered in this paper over a range of smoothing lengths. Although this cannot be considered to be a serious quantitative test of the cosmic string paradigm for structure formation, it provides a good motivation for arguing that some statistical features of topological defect models of structure formation are better than random phase models predicted by most inflationary scenarios.

## Acknowledgements

PPA is supported by JNICT. I thank my supervisor E.P.S. Shellard and Robert Caldwell for helpful comments on this manuscript.

## Appendix 1: Generation of perturbations for the toy models

In the sphere model a number $N \pm \sqrt{N}$ of spheres were put, per horizon per Hubble time, at random positions in the box. For each sphere we chose a position **p** at random. Particles were then displaced according to the Zel'dovich approximation. Consider a sphere laid down at an instant t. A particle at a position **p'** would move from that time $t$ till the present time $t_0$ a comoving distance given (in linear theory) by

$$\vec{\psi} = \frac{3}{2}GMt_i^2 \frac{\mathbf{x}}{|\mathbf{x}|^3} b_s(t), \tag{45}$$

where $\mathbf{x} = \mathbf{p}' - \mathbf{p}$ and

$$b_s(t) = \left(1 - \frac{2}{5}\frac{t}{t_0} - \frac{3}{5}\left(\frac{t_0}{t}\right)^{2/3}\right). \tag{46}$$

If we account for the growth of the mass of the loops chopped off by the network of cosmic strings ($M \propto t$) we have that the quotient of the perturbations laid down at two different times $t_1$ and $t_2$ at the same position in space is given by

$$\frac{\vec{\psi}(t_1)}{\vec{\psi}(t_2)} = \frac{t_2 \times b_s(t_1)}{t_1 \times b_s(t_2)}. \tag{47}$$

We can see that in the case of the sphere model (only) perturbations seeded at later times can have larger amplitude.

In the filament model for each filament a point **p** and a unit vector **v** were chosen at random. Particles were then displaced according to the Zel'dovich approximation. Let us consider a particle at a position $\mathbf{p}'$ and let us define the vector $\mathbf{y} = \mathbf{p}' - \mathbf{p}$. Consider the vector $\mathbf{x} = \mathbf{y} - (\mathbf{y} \cdot \mathbf{v})\mathbf{v}$ and assume that the filament has a size given by $S_f$. If $|(\mathbf{y} \cdot \mathbf{v})\mathbf{v}| < 0.5 S_f$ then a particle at a position $\mathbf{p}'$ would move a comoving distance given by

$$\vec{\psi} = -\frac{6}{5} G M_L t_i^2 \frac{\mathbf{x}}{|\mathbf{x}|^2} b_f(t). \tag{48}$$

where

$$b_f(t) = \left( \left(\frac{t_0}{t}\right)^{2/3} \ln\left(\frac{t_0}{t}\right) - \frac{3}{5}\left(\frac{t_0}{t}\right)^{2/3} + \frac{3}{5}\frac{t}{t_0} \right). \tag{49}$$

If $|(\mathbf{y} \cdot \mathbf{v})\mathbf{v}| > 0.5 S_f$ the particle would not move at all. The mass per unit length of the cosmic strings is approximately constant over time ($M_L$ = const) and so we have that the quotient of the perturbations laid down at two different times $t_1$ and $t_2$ at the same position in space is given by

$$\frac{\vec{\psi}(t_1)}{\vec{\psi}(t_2)} = \frac{b_f(t_1)}{b_f(t_2)}. \tag{50}$$

Consequently, the perturbations seeded at earlier times will have larger amplitude. The initial filament comoving size (at $t_{eq}$) was taken to be $S_f = \lambda_{eq}$ (where $\lambda_{eq} = c t_{eq}$) and its size increases with time proportionally to the horizon so that $S_f \propto t^{1/3}$. So, at later times larger structures are formed but those will be less dense.

In the wake model for each filament a point $\mathbf{p}$ and a unit vector $\mathbf{v}$ were also chosen at random. Particles were then displaced according to the Zel'dovich approximation. Let us consider the vector $\mathbf{A} = \begin{pmatrix} 0 \\ 0 \\ 1 \end{pmatrix}$ and a vector perpendicular to it $\mathbf{B} = \begin{pmatrix} \sin\alpha \\ \cos\alpha \\ 0 \end{pmatrix}$. Consider the rotation matrices

$$\mathbf{M2} = \begin{pmatrix} 1 & 0 & 0 \\ 0 & \cos\theta & \sin\theta \\ 0 & -\sin\theta & \cos\theta \end{pmatrix}, \tag{51}$$

$$\mathbf{M1} = \begin{pmatrix} \cos\beta & 0 & \sin\beta \\ 0 & 1 & 0 \\ -\sin\beta & 0 & \cos\beta \end{pmatrix}. \tag{52}$$

The vector $\mathbf{A}' = \mathbf{M2} \cdot \mathbf{M1} \cdot \mathbf{A}$ obtained as the result of multiplying the matrices $\mathbf{M1}$ and $\mathbf{M2}$ by $\mathbf{A}$ is a new vector given by $\mathbf{A}' = (\sin\beta, \sin\theta\cos\beta, \cos\theta\cos\beta)$. The vector $\mathbf{B}' = \mathbf{M2} \cdot \mathbf{M1} \cdot \mathbf{B}$ is perpendicular to $\mathbf{A}'$ and is given by

$$\mathbf{B}' = \begin{pmatrix} \cos\beta \sin\alpha \\ \cos\theta \cos\alpha - \sin\theta \sin\beta \sin\alpha \\ -\sin\theta \cos\alpha - \cos\theta \sin\beta \sin\alpha \end{pmatrix}. \tag{53}$$

The angles $\beta$ and $\theta$ were chosen subject to the constraint $\mathbf{v} = \mathbf{A'}$ and $\alpha$ was choosen at random in the interval $0 \leq \alpha < 2\pi$. We have now two perpendicular vectors $\mathbf{A'}$ and $\mathbf{B'}$ and we need to find $\mathbf{C'}$ perpendicular to this two such that

$$\mathbf{C'}.\mathbf{A'} = 0, \tag{54}$$

$$\mathbf{C'}.\mathbf{B'} = 0, \tag{55}$$

and

$$|\mathbf{C'}| = 1. \tag{56}$$

To see how a particle at a position $\mathbf{p'}$ will move let us consider the vector $\mathbf{y} = \mathbf{p'} - \mathbf{p}$ and the vectors $\mathbf{x_1} = \mathbf{y} \cdot \mathbf{B'}$, $\mathbf{x_2} = \mathbf{y} \cdot \mathbf{C'}$ and $\mathbf{x} = \mathbf{y} - (\mathbf{y} \cdot \mathbf{x_1})\mathbf{x_1} - (\mathbf{y} \cdot \mathbf{x_2})\mathbf{x_2}$. Let us assume the size of the wake to be $S_w$. If $|\mathbf{x_1}| < 0.5 S_w$ and $|\mathbf{x_2}| < 0.5 S_w$ then a particle at a position $\mathbf{p'}$ would move a comoving distance given by

$$\vec{\psi} = -\frac{2}{5} u_i t_i \frac{\mathbf{x}}{|\mathbf{x}|} b_w(t). \tag{57}$$

where

$$b_w(t) = \left( \left(\frac{t}{t_i}\right)^{2/3} - \frac{t_i}{t} \right). \tag{58}$$

If $|\mathbf{x_1}| > 0.5 S_w$ or $|\mathbf{x_2}| > 0.5 S_w$ the particle would not move at all. The quotient of the perturbations laid down at two different times $t_1$ and $t_2$ at the same position in space is given by

$$\frac{\vec{\psi}(t_1)}{\vec{\psi}(t_2)} = \frac{b_f(t_1) \times t_2^{1/3}}{b_f(t_2) \times t_1^{1/3}}. \tag{59}$$

We can see again that the perturbations seeded at earlier times will have larger amplitude. The initial wake size was taken to be $\frac{2}{3}\lambda_{\text{eq}}$ and it grows proportionally to the horizon as in the filament case.

## Appendix 2: The simulation boxes

As we have said before the size of the simulation boxes was chosen in a way to enable direct comparison with the results from Vogeley et al. (1994) for the CfA survey. To ensure that the topology was not dominated by shot noise the smoothing length must be larger than the average intergalaxy (or interparticle) separation. To determine the maximum distance appropriate for a given choice of smoothing length Vogeley et al. (1994) found $r_{\mathrm{max}}$ such that

$$n(r_{\mathrm{max}}) = \lambda_{\mathrm{e}}^{-3}. \tag{60}$$

| $\lambda_{\mathbf{e}}(h^{-1}\mathrm{Mpc})$ | $V_{\mathrm{survey}}(h^{-1}\mathrm{Mpc})^3$ | $N_{\mathrm{res}}$ | $N_{\mathrm{galaxies}}$ |
|---|---|---|---|
| 6 | $3.31 \times 10^5$ | 260 | 5546 |
| 8 | $5.78 \times 10^5$ | 202 | 7257 |
| 10 | $8.38 \times 10^5$ | 150 | 8139 |
| 12 | $1.07 \times 10^6$ | 111 | 7777 |
| 16 | $1.51 \times 10^6$ | 66 | 8234 |
| 20 | $1.87 \times 10^6$ | 42 | 8404 |

**Table 1:** Volume statistics for the CfA survey (note that the total volume and consequently the number of resolution elements for the toy model simulations is the same as for the CfA survey).

This means that the number of galaxies included is an increasing function of the smoothing length. Table 1 shows the volume of the survey as a function of the smoothing length. The number of resolution elements, defined by $N_{\mathrm{res}} = \frac{V_{\mathrm{survey}}}{\pi^{3/2} \lambda_e^3}$, and the number of galaxies included in the topological analysis of the CfA survey are also given as a function of the smoothing length. The volume of the simulation boxes and so the number of resolution elements were chosen to be the same as for the CfA survey analysis so that direct comparison between our results and observations was possible. The smoothing length was always more than 2 times larger than the average interparticle spacing.

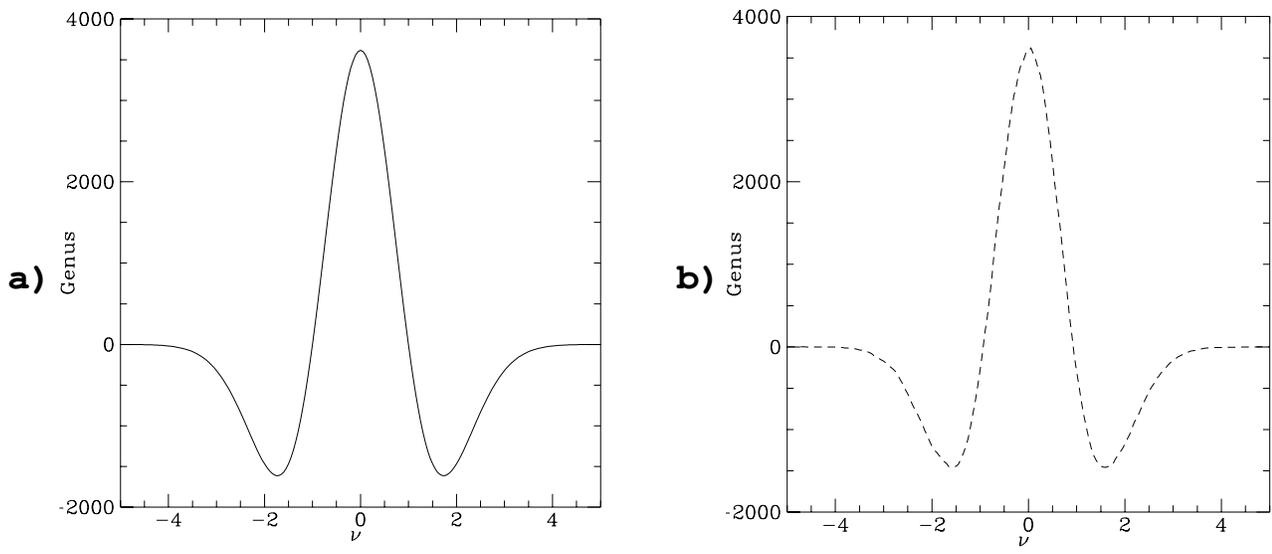

**Fig. 1** Comparison of the genus curve obtained for a $64^3$ simulation of a $P(k) \propto k$ power spectrum smoothed on a scale $\lambda_e$ equal to $\sqrt{2}$ times the grid spacing, obtained (a) analytically (b) numerically using our program to calculate the genus.

**Fig. 2** Isodensity contours with $\delta = 1.5\nu$ for (a) the 5 filament model b) the 5 wake model c) the 25 sphere model with cold dark matter. The box size is $3.13 \times 10^5 h^{-1} \mathrm{Mpc}^3$, and the smoothing length is $\lambda_\mathrm{e} = 6h^{-1}\mathrm{Mpc}$.

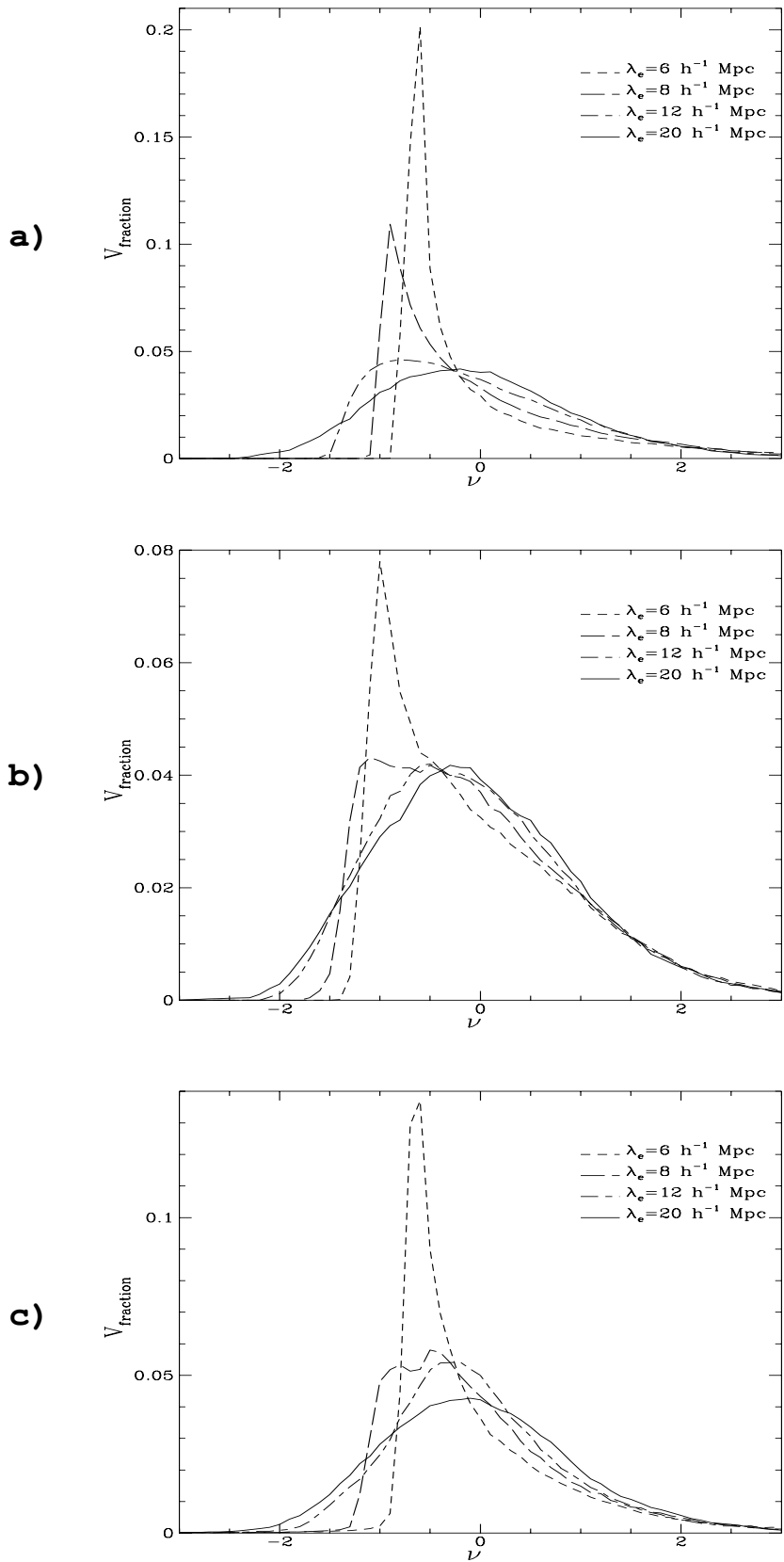

**Fig. 3** Density probability distribution for several models studied with cold dark matter as a function of $\nu$ calculated directly from the variance of the density distribution (a) 5 filament model (b) 5 wake model and (c) 25 sphere model.

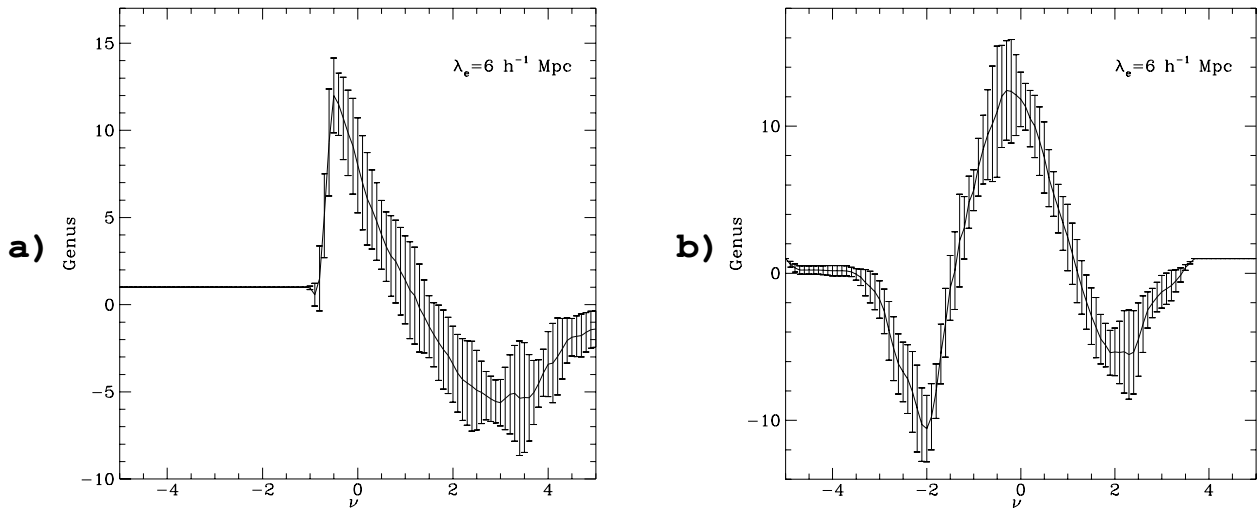

**Fig. 4** Comparison of genus curves for the 5 filament model assuming the dark matter to be cold. (a) using the volume fraction to prametrize $\nu$ (b) Calculating $\nu$ directly from the variance of the density distribution.

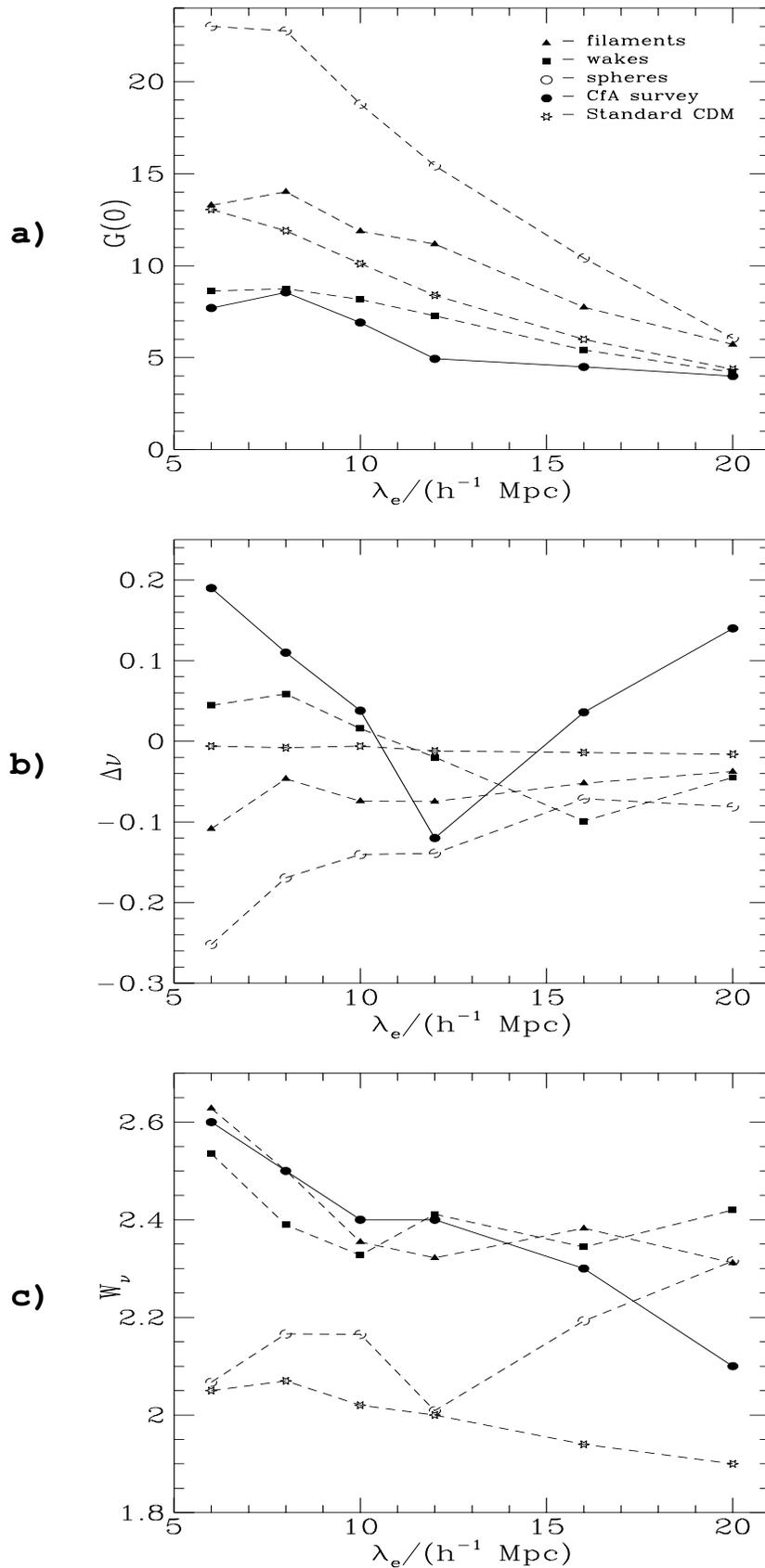

**Fig. 5** Comparison of genus curves for the 5 wake, 5 filament and 25 sphere models with both standard CDM genus curves and genus curves obtained from the CfA survey. Error bars are not included for clarity (see fig 8). It is clear from the picture that the most favoured model is the 5 wake model.

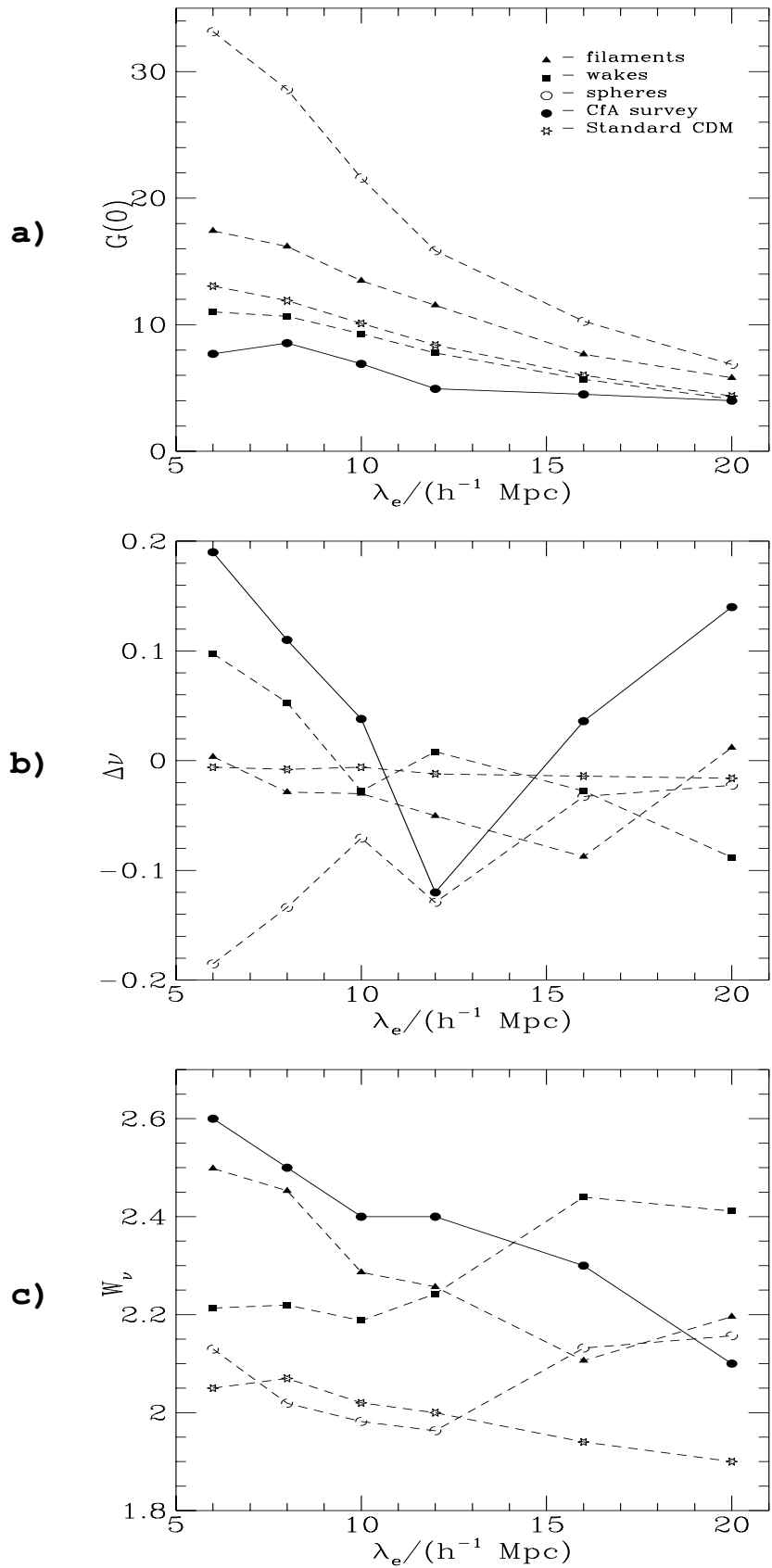

**Fig. 6** Comparison of genus curves for the 10 wakes, 10 filaments and 50 spheres models with both standard CDM genus curves and genus curves obtained from the CfA survey.

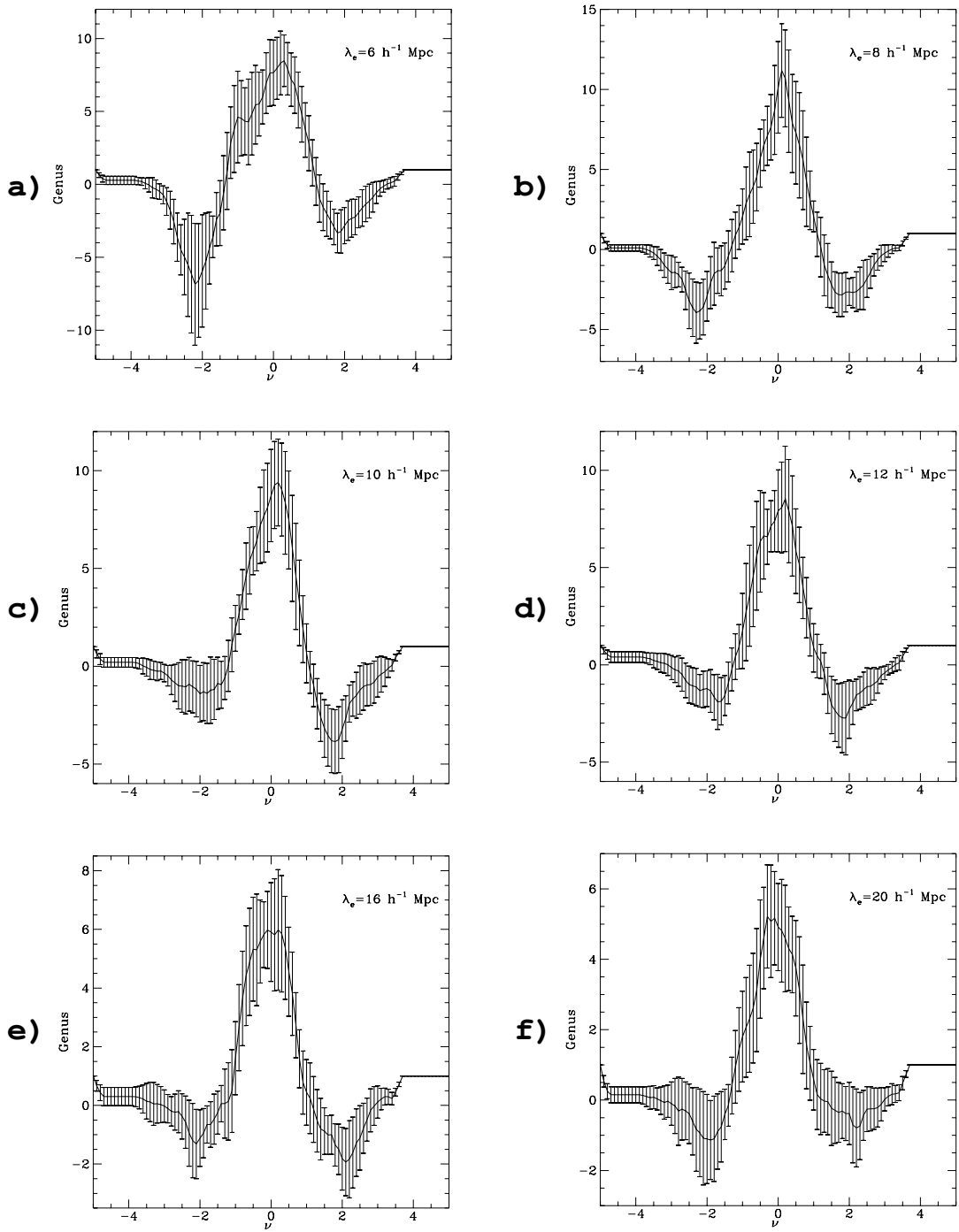

**Fig. 7** Genus curves for the 5 wake model. Line represents the average genus curve among 10 realizations of the model. Error bars are one-sigma.

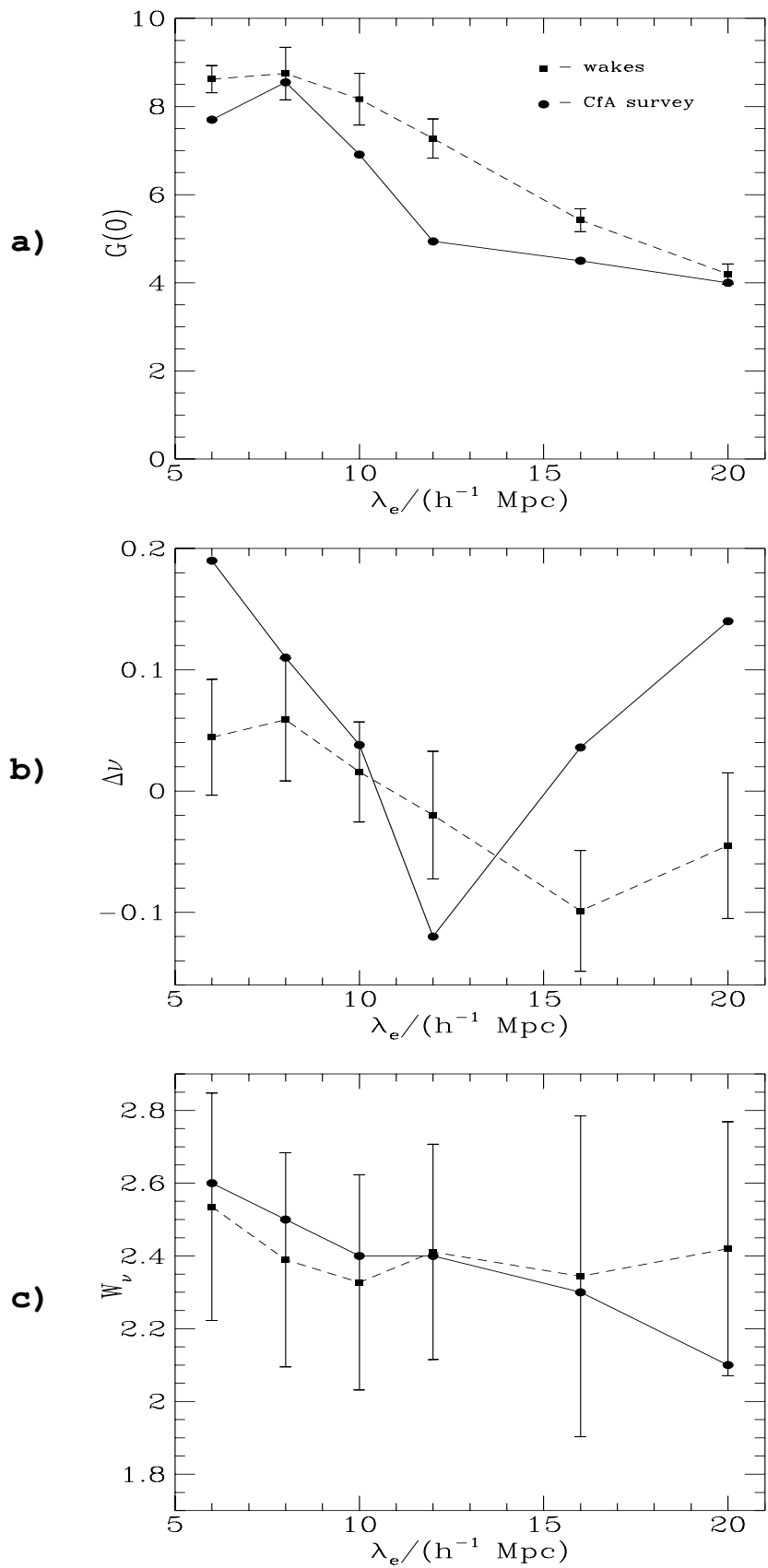

**Fig. 8** Statistical comparison of genus curves obtained from the CfA survey with genus curves obtained from 10 realizations of the 5 wake model. Error bars on the 5 wakes model are one-sigma.

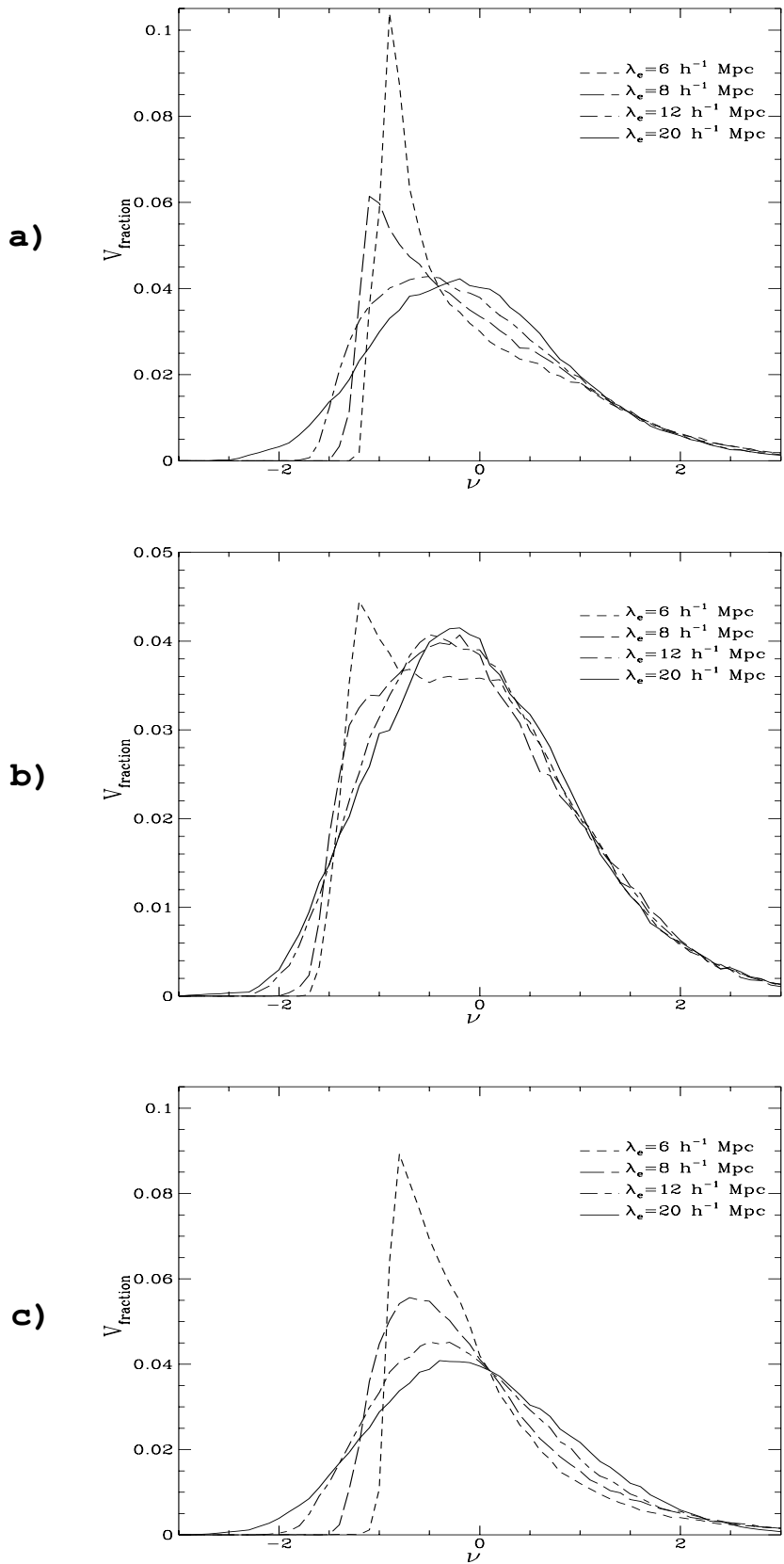

**Fig. 9** Density probability distribution for several models studied with hot dark matter as a function of $\nu$ calculated directly from the density distribution (a) 5 filament model (b) 5 wake model and (c) 25 sphere model.

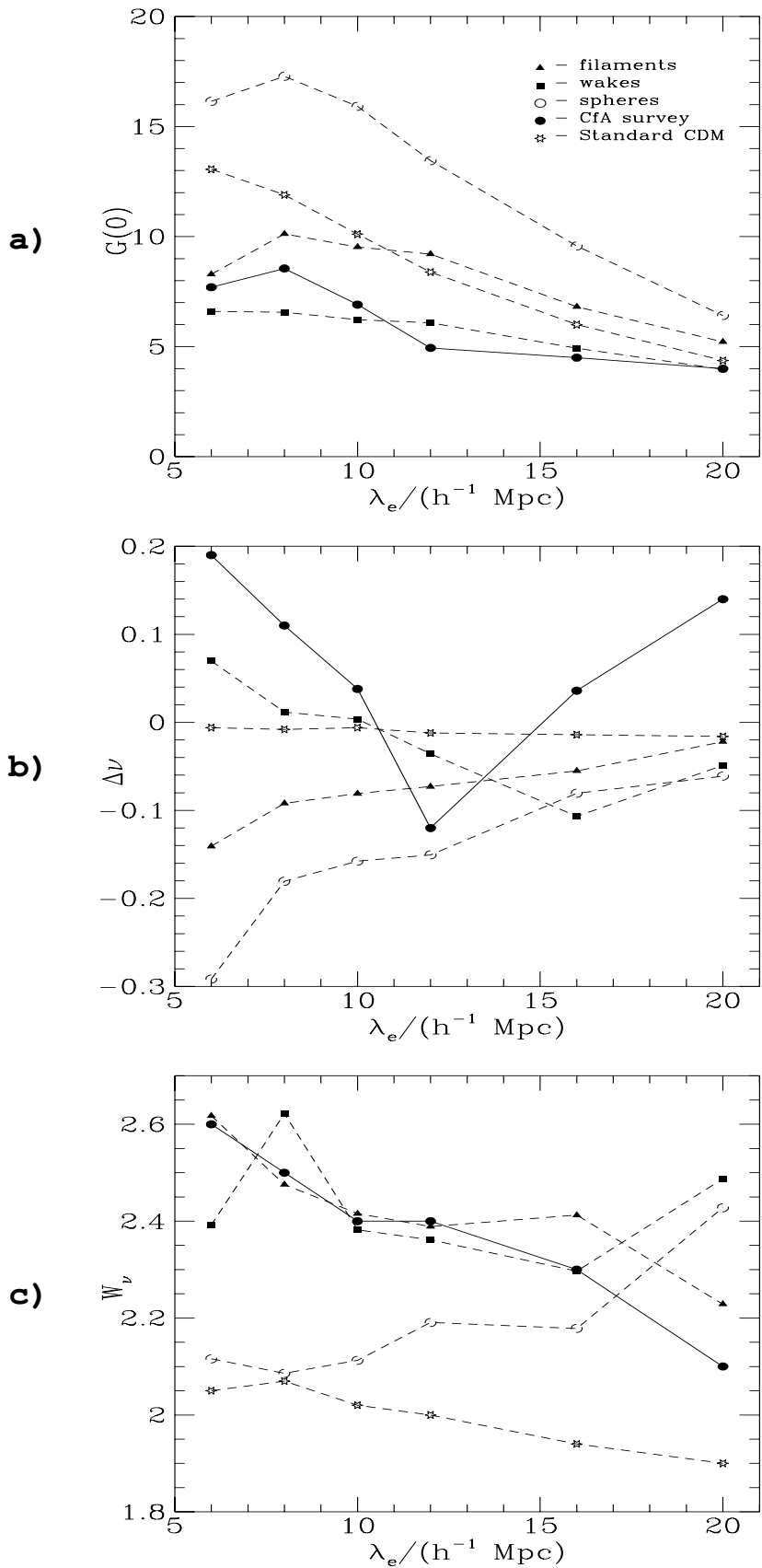

**Fig. 10** Comparison of genus curves for the 5 wake, 5 filament and 25 sphere models in the context of hot dark matter with both standard CDM genus curves and genus curves obtained from the CfA survey. Error bars are not included for clarity (see fig 13).

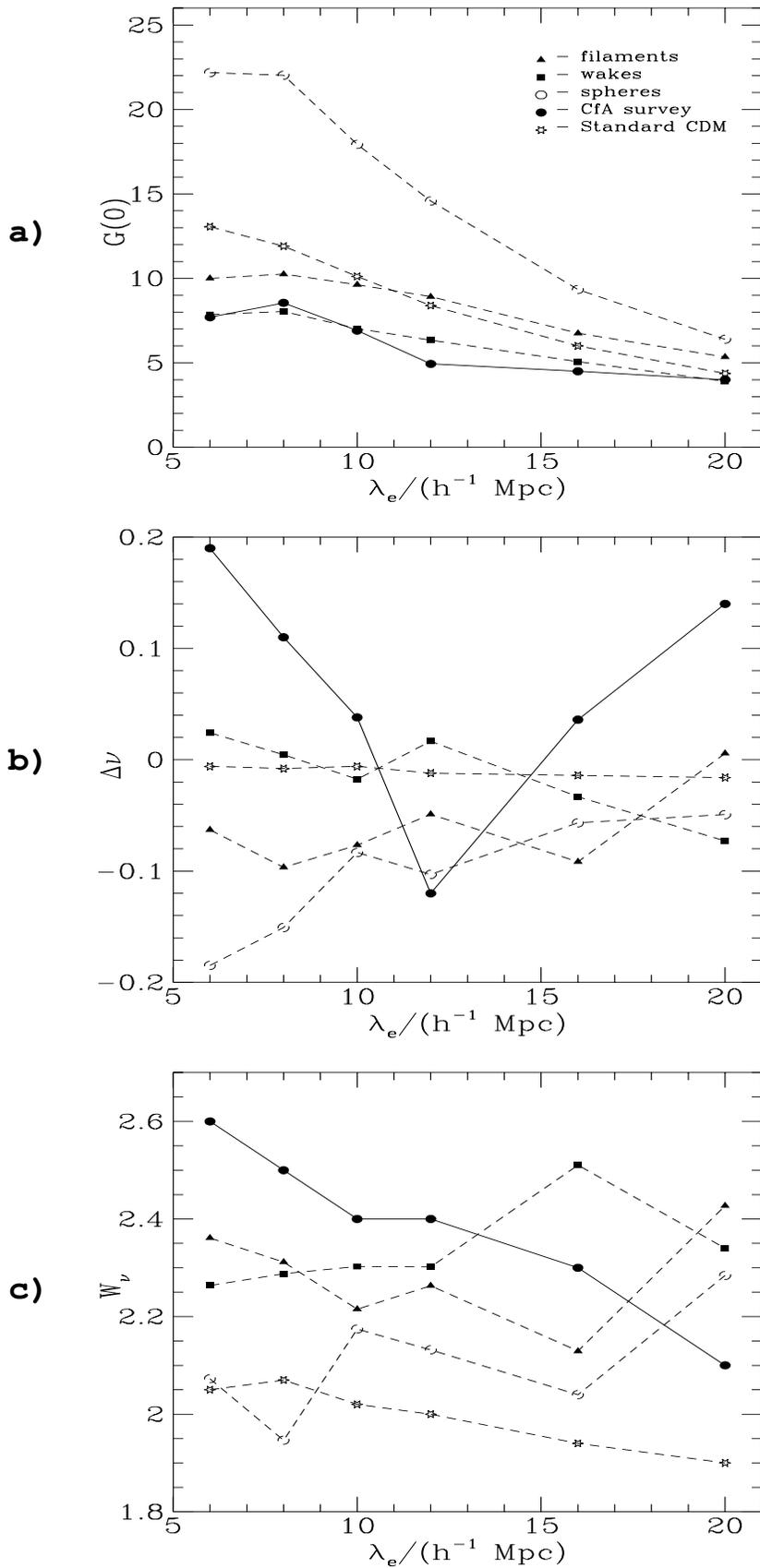

**Fig. 11** Comparison of genus curves for the 10 wake, 10 filament and 50 sphere models in the context of hot dark matter with both standard CDM genus curves and genus curves obtained from the CfA survey. It is clear from the picture that the most favoured model is the 10 wake model.

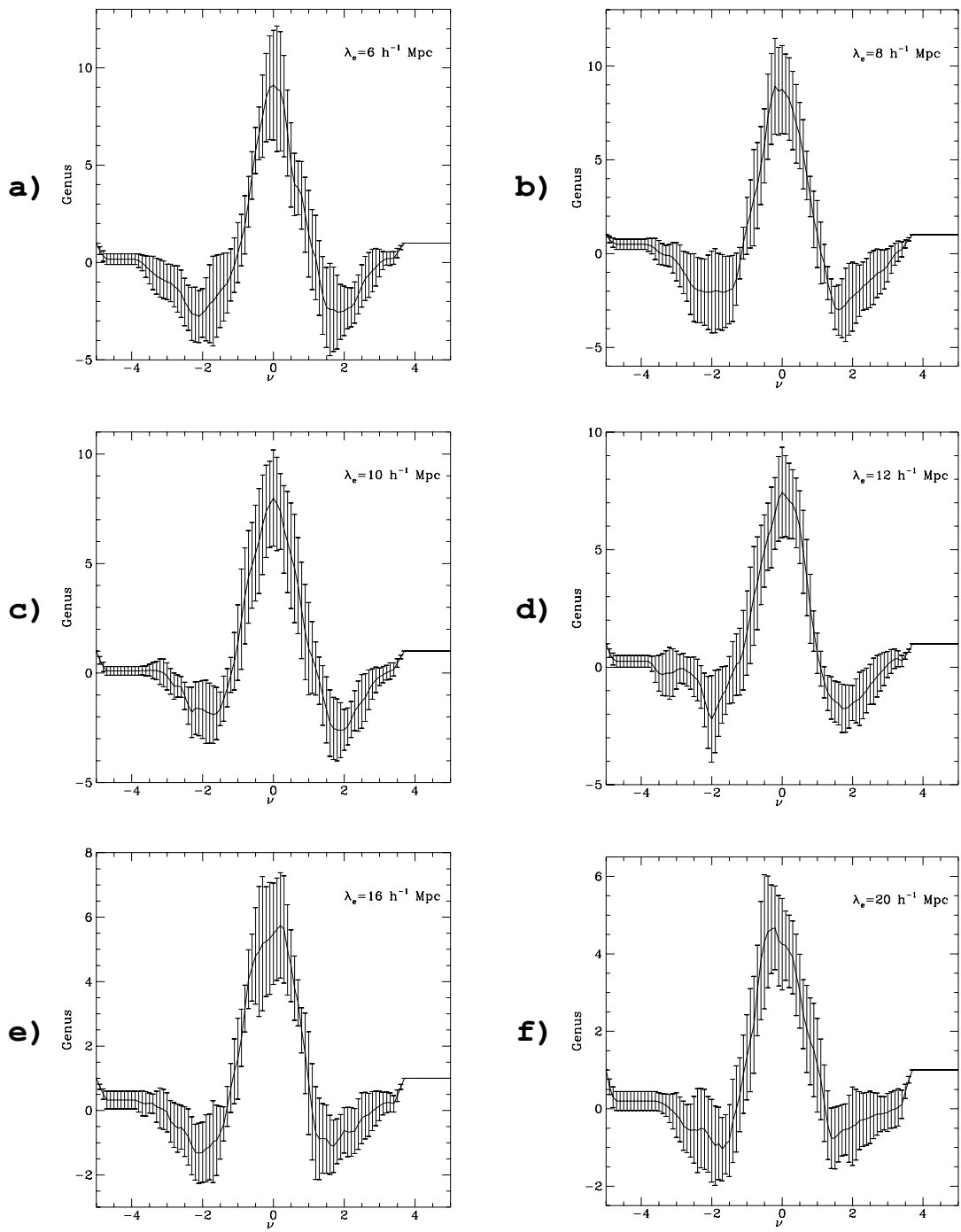

**Fig. 12** Genus curves for the 10 wake model with HDM. Line represents the average genus curve among 10 realizations of the model. Error bars are one-sigma.

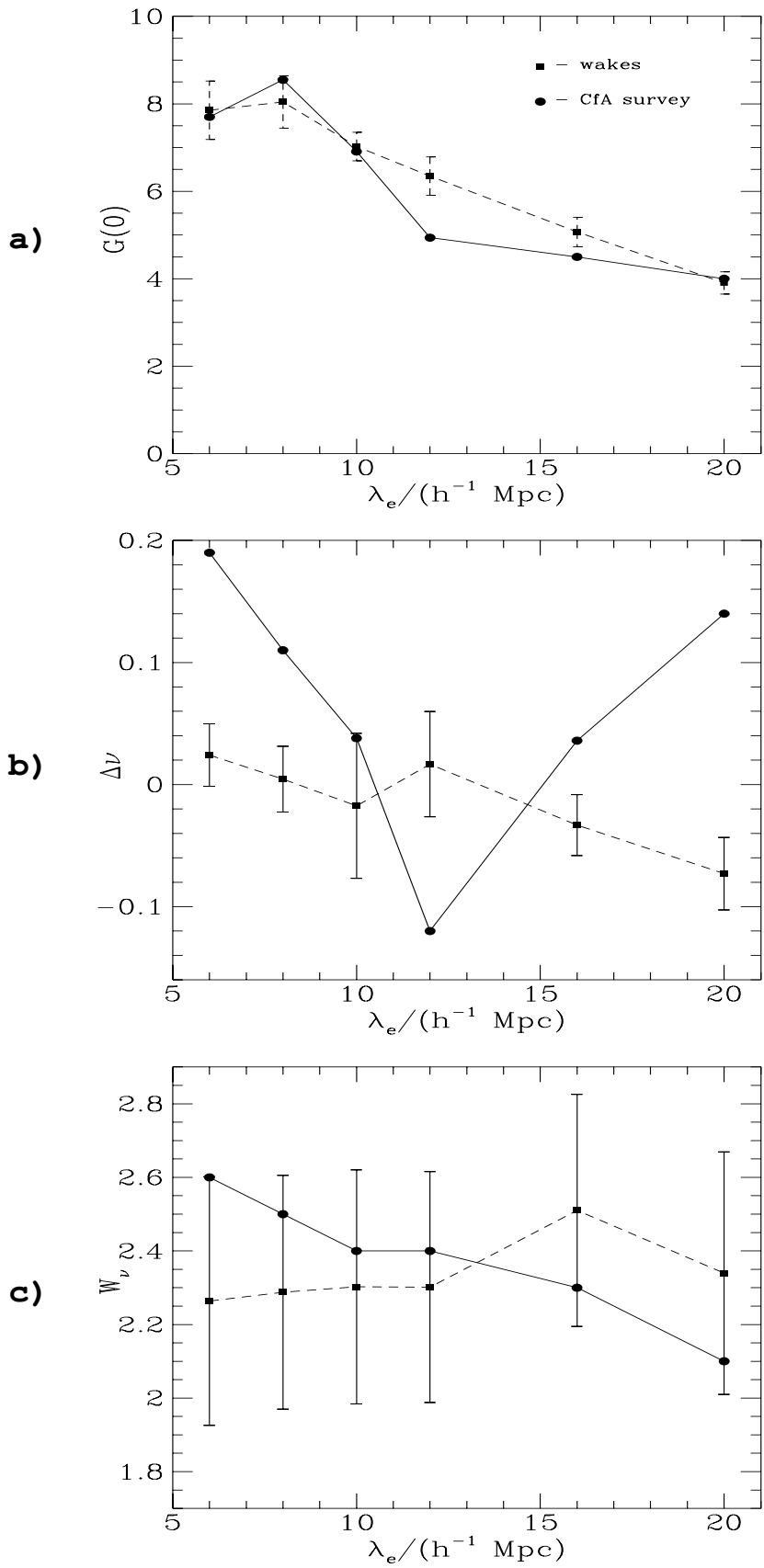

**Fig. 13** Statistical comparison of genus curves obtained from the CfA survey with genus curves obtained from 10 realizations of the 10 wake model with HDM. Error bars are one-sigma